\documentclass[a4paper]{article}

\usepackage{INTERSPEECH2022}
\usepackage{multirow}
\usepackage{hyperref}
\title{Speech Emotion: Investigating Model Representations, Multi-Task Learning and Knowledge Distillation}
\name{Vikramjit Mitra, Hsiang-Yun Sherry Chien, Vasudha Kowtha, Joseph Yitan Cheng, Erdrin Azemi}
\address{Apple}
\email{\{vmitra, sherry.chien, vkowtha, jycheng, erdrin\}@apple.com}

\begin{document}

\maketitle
\begin{abstract}
Estimating dimensional emotions, such as activation, valence and dominance, from acoustic speech signals has been widely explored over the past few years. While accurate estimation of activation and dominance from speech seem to be possible, the same for valence remains challenging. Previous research has shown that the use of lexical information can improve valence estimation performance. Lexical information can be obtained from pre-trained acoustic models, where the learned representations can improve valence estimation from speech. We investigate the use of pre-trained model representations to improve valence estimation from acoustic speech signal. We also explore fusion of representations to improve emotion estimation across all three emotion dimensions: activation, valence and dominance. Additionally, we investigate if representations from pre-trained models can be distilled into models trained with low-level features, resulting in models with a less number of parameters. We show that fusion of pre-trained model embeddings result in a ${79\%}$ relative improvement in concordance correlation coefficient ($CCC$) on valence estimation compared to standard acoustic feature baseline (mel-filterbank energies), while distillation from pre-trained model embeddings to lower-dimensional representations yielded a relative ${12\%}$ improvement. Such performance gains were observed over two evaluation sets, indicating that our proposed architecture generalizes across those evaluation sets. We report new state-of-the-art "text-free" acoustic-only dimensional emotion estimation $CCC$ values on two MSP-Podcast evaluation sets. 

\end{abstract}

\section{Introduction}

Human speech communication broadly consists of two layers: the linguistic layer, which conveys messages in the form of words and their meaning, and the paralinguistic layer, which conveys how those words have been said, including vocal expressiveness or emotional tone. While computational models for recognizing words from speech have been exhaustively explored \cite{Rabiner89-ATO, seide2011conversational, hannun2014first, mitra2015time}, acoustic-based speech emotion recognition modeling has only recently garnered attention. 

Computational models for speech emotion recognition have explored two directions: (a) basic discrete emotions, such as happy, sad, angry etc., and (b) dimensional emotions, such as levels of activation, valence and dominance. Discrete emotion based models suffer from a lack of a list of standard emotions, where the choice can vary from Ekman's six basic emotions \cite{ekman1992argument} (e.g., happy, sad, anger, disgust, fear and surprise) to the more comprehensive twenty-seven emotion categories as proposed by Cowen and Keltner \cite{cowen2017self}. Variation in discrete emotion definitions may result in annotation complexities, difficulty in realizing datasets with consistent discrete emotion labels, and fail to include rare emotional states such as boredom. Dimensional emotions vary along three principal affect dimensions: activation, valence, and dominance \cite{posner2005circumplex}, where an emotion can be interpreted as a point within these three dimensions. In dimensional emotion, activation reflects the energy in voice, valence indicates the negative versus positive emotion, and dominance specifies how strong or submissive/meek one may sound.

Early studies on speech-based emotion detection have focused on acted or elicited emotions \cite{busso2008iemocap, jackson2014surrey}, where actors were recorded while speaking with specified emotions. Unfortunately, models trained with elicited emotions often fail to generalize for spontaneous subtle emotions \cite{douglas2005multimodal}. More recently, attention has been given to datasets with spontaneous emotions \cite{mariooryad2014building}. However, obtaining ground-truth emotion labels for spontaneous speech datasets is challenging, because it often results in varying degrees of grader agreement and/or grading quality.

Initial speech emotion recognition models rely heavily on acoustic features that emphasize expert-level knowledge \cite{schuller2012avec, eyben2015geneva}, however such representations are often high dimensional. Model driven approaches have demonstrated that simple acoustic features, such as mel-filterbank (MFB) features, can generate emotion recognition performance as-good-as and often better than high dimensional acoustic features \cite{khalil2019speech, mitra2019leveraging, kowtha2020detecting}. However, while estimating activation and dominance from simple acoustic features seems to be promising, estimating valence from such low-level features remains a challenging task. It has been observed that valence seems to be strongly correlated with lexical information \cite{aldeneh2017pooling}, whereas acoustic-based information is better correlated with activation and dominance \cite{atmaja2020dimensional}. 

Studies have shown that combining text- and acoustic-based representations can boost model performance on emotion recognition from speech \cite{sahu2019multi, ghriss2022sentiment, srinivasan2021representation, siriwardhana2020jointly}. In addition, recent work demonstrated that using representations from pre-trained acoustic models can improve emotion recognition performance \cite{siriwardhana2020jointly, pepino2021emotion}. More interestingly, such pre-trained features can substantially improve valence estimation \cite{srinivasan2021representation}. It is hypothesized that given the self-supervised learning architecture of the pre-trained models and the large speech data-sets they were exposed to, representations generated by these pre-trained models may contain lexical information which facilitate better valence estimation \cite{srinivasan2021representation}. However, pre-trained representations are often high dimensional which may render the resulting emotion models to be complex both in terms of memory and computation time. It is thus desirable to leverage the performance of a smaller acoustic model by distilling knowledge from complex models \cite{chebotar2016distilling}.  

In this work we used a spontaneous speech corpus labelled with dimensional emotions and investigate the following hypotheses:
(1) If features learned from pre-trained acoustic models can improve dimensional emotion estimation by comparing its performance with traditional low-level acoustic features.
(2) If fusion of features from multiple pre-trained models can further improve model performance on dimensional emotion estimation.
(3) If knowledge from high-dimensional, complex models can be transferred to less complex models.

We demonstrate that  
(1) Features from pre-trained models significantly improve valence estimation.
(2) The best result of valence estimation could be obtained by fusion of features from multiple pre-trained models (${\approx 79\%}$ relative performance gain).
(3) Distillation of information from a complex speech emotion model trained on high-dimensional features can significantly enhance the performance of a smaller model trained on MFB features with a ${\approx 12\%}$ relative performance gain on valence estimation.

Note that unlike the work presented in \cite{ghriss2022sentiment, srinivasan2021representation, siriwardhana2020jointly} where text information was part of the input to the models, the emotion models presented in this paper are completely \textbf{"text free"}. While certain pre-trained models adapted in the current study (e.g., models that were fine-tuned to ASR task such as HuBERT ASR\cite{hsu2021hubert}) were exposed to text information, we have only extracted the embeddings from those models without explicitly using any text based representation in any of our experiments.

The rest of the paper is organized as follows: Section (2) presents the dataset used in our study, Section (3) introduces feature representations investigated and details on the acoustic model and its parameters, Section (4) presents the results, followed by conclusions in Section (5).

\section{Data}

In this work we use the MSP-Podcast dataset 1.6 \cite{mariooryad2014building, lotfian2017building} that contains speech spoken by English speakers collected from online audio shows, covering topics such as politics, sports, entertainment, etc. The speech segments in this dataset contain single speaker utterances with duration between 2.75 and 11 seconds. The dataset contains ${\approx}$ 85 hours of speech and it comes with activation, valence and dominance scores from multiple graders specified on a 7-point likert scale. The dataset is partitioned into training, validation, and testing splits. To make our results comparable to literature, we report results on MSP-Podcast 1.3 and 1.6 (MSP-eval\_1.3 and MSP-eval\_1.6).

\section{Methods}
\subsection{Acoustic Features}
The baseline acoustic feature consists of 40-dimensional MFB energies, which were analyzed at a 25ms window at a frame interval of 20ms. The MFB features are appended with pitch, pitch-delta and voicing features, resulting in a 43-dimensional acoustic feature, denoted as ${MFB+F_0}$ feature. 
\vspace{-1mm}
\subsection{Features from Pre-Trained Models}
We explored embeddings generated from two pre-trained acoustic models, WAV2VEC2.0 \cite{baevski2020wav2vec} and HuBERT \cite{hsu2021hubert}. Both pre-trained models contain a feature encoder which extracts acoustic features through temporal convolutions, and a "contextualizer" which contains multi-layer transformer encoders for learning contextualized representations by reconstructing masked target features. The reconstruction task requires the model to capture contextual relationships of the speech signal which is useful for various downstream tasks. There are two major differences between WAV2VEC2.0 and HuBERT: First, the target features in WAV2VEC2.0 are generated by quantizing latent features, while the target features in HuBERT, are generated from clustering the MFCC features. Second, WAV2VEC2.0 learns embeddings by minimizing contrastive loss, where the contextualizer needs to identify the true quantized latent feature in a set of candidate features including sampled distractors; while the contextualizer in HuBERT only focused on reconstructing the masked part of the signals. With the BERT-like architecture, HuBERT was able to learn a combined acoustic and language model and outperformed WAV2VEC2.0 on downstream ASR tasks \cite{hsu2021hubert}. 

In our study, we extracted embeddings from both base and large pre-trained models. The base models were pre-trained on 960 hours of speech from \textit{Librispeech} dataset with 12 transformer layers and 768 embedding dimensions, while the large models were pre-trained on 60k hours of speech from \textit{Libri-Light} dataset with 24 transformer layers and 1024 embedding dimensions. We also extracted embeddings from pre-trained models that were fine-tuned to ASR tasks with 100 hours and 960 hours of speech from \textit{Librispeech} dataset, where we expect the embeddings to capture more lexical information and further improve valence decoding \footnote{Embeddings were extracted using pre-trained models and pipelines provided in torchaudio: \url{https://pytorch.org/audio/stable/pipelines}}.  
\vspace{-1mm}
\subsection{Dimensional Emotion Estimation System}
\begin{figure}[htb]
\begin{minipage}[b]{1.0\linewidth}
  \centering
  \centerline{\includegraphics[width=8.0cm]{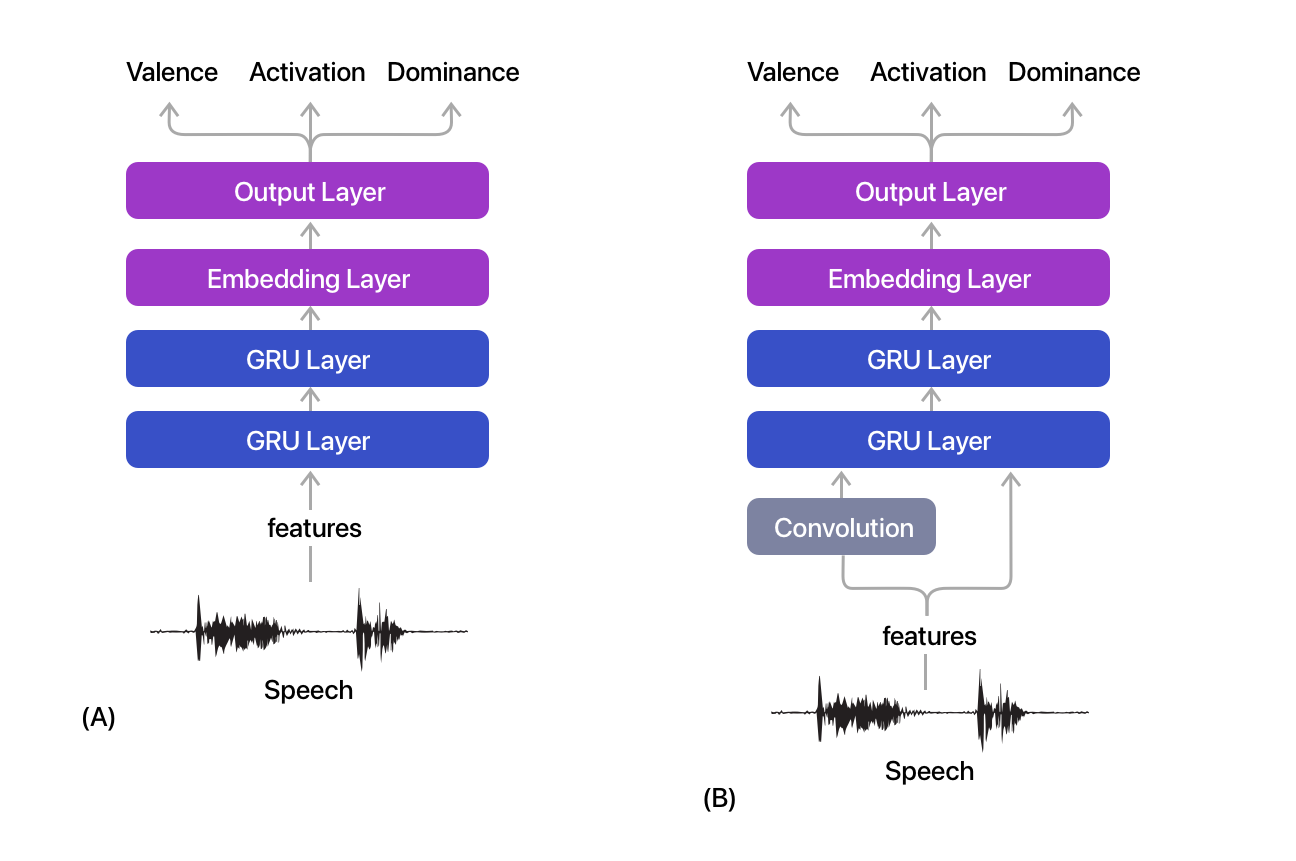}}
\end{minipage}
\caption{Architecture of the multi-task (A) GRU network and (B) Time convolution GRU network (TCGRU)}
\label{fig:fig1}
\end{figure}

We used a two-layer Gated Recurrent Unit (GRU) network consisting of 128 neurons in the recurrent and the embedding layers, shown in Figure \ref{fig:fig1} (A), to train the baseline emotion (regression) model. In addition, we explored a time convolutional (TC) layer where the number of convolutional layers (with filter size = 3) was the same as the number of input feature dimensions. The TC layer had skip connections, where acoustic features were also fed directly to the GRU layer as shown in Figure \ref{fig:fig1} (B) as the TCGRU network. The input to the model was the $MFB+F_0$ features, and the output was 3 dimensional emotions: activation, valence, and dominance. The model was tuned using a held-out validation-set (specified by the MSP 1.6 data). Concordance correlation coefficient ($CCC$) is used as the loss function ($L_{ccc}$), as shown in equation (1), where ${L_{ccc}}$ is a combination ($\alpha=1/3$ and $\beta=1/3$) of CCC's obtained from each of the activation, valence, and dominance dimensions. $CCC$ for each dimension is defined by equation (2), where ${\mu _{x}}$ and ${\mu _{y}}$ are the means, ${\sigma _{x}^{2}}$ and ${\sigma _{y}^{2}}$ are the corresponding variances for the estimated and ground-truth variables, and ${\rho}$ is the correlation coefficient between those two variables. The models were trained with a mini-batch size of 32, and using Adam optimizer with a learning rate of 0.0005. For all model training steps, early stoppage was performed based on $CCC$ values from the validation-set. We observed that multi-task learning to be beneficial, where we introduced an additional output classification layer for recognizing seven discrete emotion labels provided by the MSP 1.6 dataset. The cross-entropy loss from the classification layer is added as an auxiliary task with an empirically defined weight of 0.2 to the overall loss.

\vspace{-2mm}
\begin{equation}
\begin{aligned}
{L_{ccc}:= 1 - \left (\alpha CCC_{v}+\beta CCC_{a}+(1-\alpha-\beta)CCC_{d} \right )}, \\
\end{aligned}
 \label{eq1}
\end{equation}

\vspace{-2mm}
\begin{equation}
\begin{aligned}
CCC &= \frac {2\rho \sigma_x \sigma_y}{\sigma_x^2+\sigma_y^2 +(\mu_x-\mu_y)^2 }.
\end{aligned}
 \label{eq2}
\end{equation}
\vspace{-2mm}
\vspace{-3mm}

\subsection{Feature Distillation}
Although pre-trained embeddings are expected to improve valence estimation, they are typically high dimensional (e.g., HuBERT embeddings have 1024 dimensions) which causes the resulting dimensional emotion detection models to have a larger memory footprint (1.5MB to 4.2MB in size depending upon input feature dimension and model architecture). Extracting pre-trained model embeddings also requires an extra step of running the pre-trained models on the acoustic speech signal, which is computationally expensive. On the other hand, having low-dimensional features such as MFBs can significantly reduce both the computational and memory complexities of the speech emotion models (size $\approx$ 0.2MB), but those features struggled with valence estimation (Table \ref{tab:table1}). One hypothesis is that such low level features fail to induce relevant representations within the speech emotion model that is conducive to robust valence estimation, where pre-trained model embeddings are successful owing to the self-supervised learning architecture and the large amount of training data. It is noteworthy that the pre-trained models are themselves trained with low-level acoustic features and may have learned better representations of acoustic speech, which are useful for a wide range of speech tasks \cite{mohamed2021arabic, yi2020applying, vaessen2021fine, chen2021large, srinivasan2021representation}.

To investigate if knowledge can be distilled from a complex but powerful model into a simpler model, we explored distilling the embeddings learned from pre-trained feature based emotion models into MFB feature based emotion models during the training step using a teacher-student framework. Let us assume, the embeddings from the teacher network (i.e.\ the HuBERT pre-trained feature based speech emotion model) are ${E_{T,i}}$, where the subscripts ${T}$ represents the Teacher model, and ${i}$ the speech sample, and the learned embeddings from the student model (i.e.\ the MFB feature trained speech emotion model) be ${\hat{E}_{S,i}}$. Let the ground truth labels be ${l_{k,i}}$ where ${k\in \{a,v,d\}}$ and the estimated labels from the HuBERT emotion model be ${\tilde{l}_{k,i}}$. We define the distillation loss as:
\vspace{-3mm}
\begin{equation}
  L_{dis,k} = \left(1 - \frac {1}{N}\sum_i \frac{E_{T,i}\cdot\hat{E}_{S,i}}{max \left( ||{E_{T,i}}||_2 \cdot ||\hat{E}_{S,i}||_2,\epsilon\right)} \right) \cdot \gamma_{i},
  \label{eq3}
\end{equation}
\begin{equation}
  \gamma_{i} = 1 - \frac {1}{M} \left ( \frac {\sum_k |l_{k,i} - \tilde{l}_{k,i}|}{3} \right ).
  \label{eq4}
\end{equation}

In equation \ref{eq3} the distillation loss is computed using the mean cosine distance between the teacher (${E_{T,i}}$) and the student (${\hat{E}_{S,i}}$) network embeddings (where $N$ is the embedding length), and is weighted by the residual error between the target (${l_{k,i}}$) and the estimated (${\tilde{l}_{k,i}}$) labels of the teacher network. Note that $M$ represents the dynamic range of the target labels ${l_{k,i}}$ in the dataset, and the residuals are weighted by the dynamic range $M$ to keep the values between 0 and 1. $\gamma_{i}$ provides a confidence measure for each sample point, where it is closer to 1 for samples that have low residual error from the teacher network and closer to zero when the residuals are high. Hence, $\gamma_{i}$ works as a soft coefficient that weights the distillation loss, emphasizing more on the teacher embeddings that result in lower residual error in the teacher network and de-emphasizing the ones that generate higher residual error. The final loss of the model training is defined as below - 

\vspace{-3mm}
\begin{equation}
  L_{k} = \kappa (L_{ccc,k} + L_{CE,k}) + \lambda L_{dis,k},
  \label{eq5}
\end{equation}
where $\kappa$ and $\lambda$ are coefficients used to weigh the different loss terms ($(L_{ccc,k}$ for the $CCC$ loss of the dimensional emotion, $(L_{CE,k}$ loss for the discrete emotions and the $L_{dis,k}$ for knowledge distillation) as a function of epoch. For the first 40 epochs we use $\kappa$ = 0.001 and $\lambda$ = 1 and for the remaining epochs we use $\kappa$ = 1 and $\lambda$ = 0.01. The above selection of $\kappa$ and $\lambda$ was made empirically to emphasize the distillation task initially and then emphasize on the target task toward the end of training.

\section{Results}
We first trained a GRU and a TCGRU with the baseline $MFB+F_0$ features. The models are compared against MSP-eval\_1.6 and MSP-eval\_1.3. As shown in Table~\ref{tab:table1}, TCGRU yields better ${CCC}$ across all three dimensions compared to GRU (and the difference was statistically significant with $p<0.05$). 

%lSSSSSS
\vspace{1mm}

\begin{table}[th]
%\small
\centering
\caption{Dimensional emotion estimation $CCC$ from models trained with baseline $MFB+F_0$ features}
\label{tab:table1}
  \begin{tabular}{lcccccc}
    \toprule
    \multirow{2}{*}{System} &
      \multicolumn{3}{c}{\textbf{MSP-eval\_1.3}} &
      \multicolumn{3}{c}{\textbf{MSP-eval\_1.6}} \\
      & {act} & {val} & {dom} & {act} & {val} & {dom} \\
      \midrule
    GRU & 0.71 & 0.31 & 0.65 & 0.69 & 0.31 & 0.63  \\
    TCGRU & \textbf{0.73} & \textbf{0.33} & \textbf{0.66} & \textbf{0.71} & \textbf{0.33} & \textbf{0.64} \\
    \bottomrule
  \end{tabular}
\end{table}
\vspace{-2mm}
Next, we explored using pre-trained model generated embeddings as input features to train GRU-based dimensional emotion estimation models and the results are shown in Table~\ref{tab:table2}. Table ~\ref{tab:table2} results demonstrate the strength of the pre-trained model embeddings for dimensional emotion estimation from speech compared to the baseline $MFB+F_0$ features shown in Table ~\ref{tab:table1}. We observed a substantial improvement in valence estimation performance ({$\approx 60\%$}) compared to the baseline, while improvement in activation and dominance was nominal. These findings can be attributed to the representations learned by the pre-trained models, which may contain lexical information beyond what may be captured by low-level acoustic features. 

Based on the performance of the pre-trained model embeddings in Table \ref{tab:table2}, we evaluated TCGRU model trained on HuBERT Large, HuBERT Large 100H embeddings and their combination. The results in Table \ref{tab:table3} show that TCGRU trained on fusion of the pre-trained embeddings improves estimation ($>7\%$ relative) across all three dimensional emotions, compared to the model trained with each of those two embeddings separately.

Finally, we observed that knowledge distillation from high-dimensional features can improve valence estimation from small models trained on low-dimensional acoustic features (Table \ref{tab:table4}). Note that the distilled model in Table \ref{tab:table4} failed to achieve comparable valence estimation $CCC$ as observed in the systems shown in Table \ref{tab:table3}. Low-level acoustic features ($MFB+F_0$) may be limited by their capacity to properly account for contextual information, as a consequence impacting their valence estimation performance compared to pre-trained model embedding representations.
\vspace{-2mm}
\subsection{Further Analyses on Learned Representations}
\begin{figure}[!ht]
\begin{minipage}[b]{1.0\linewidth}
  \centering
  \centerline{\includegraphics[width=6cm]{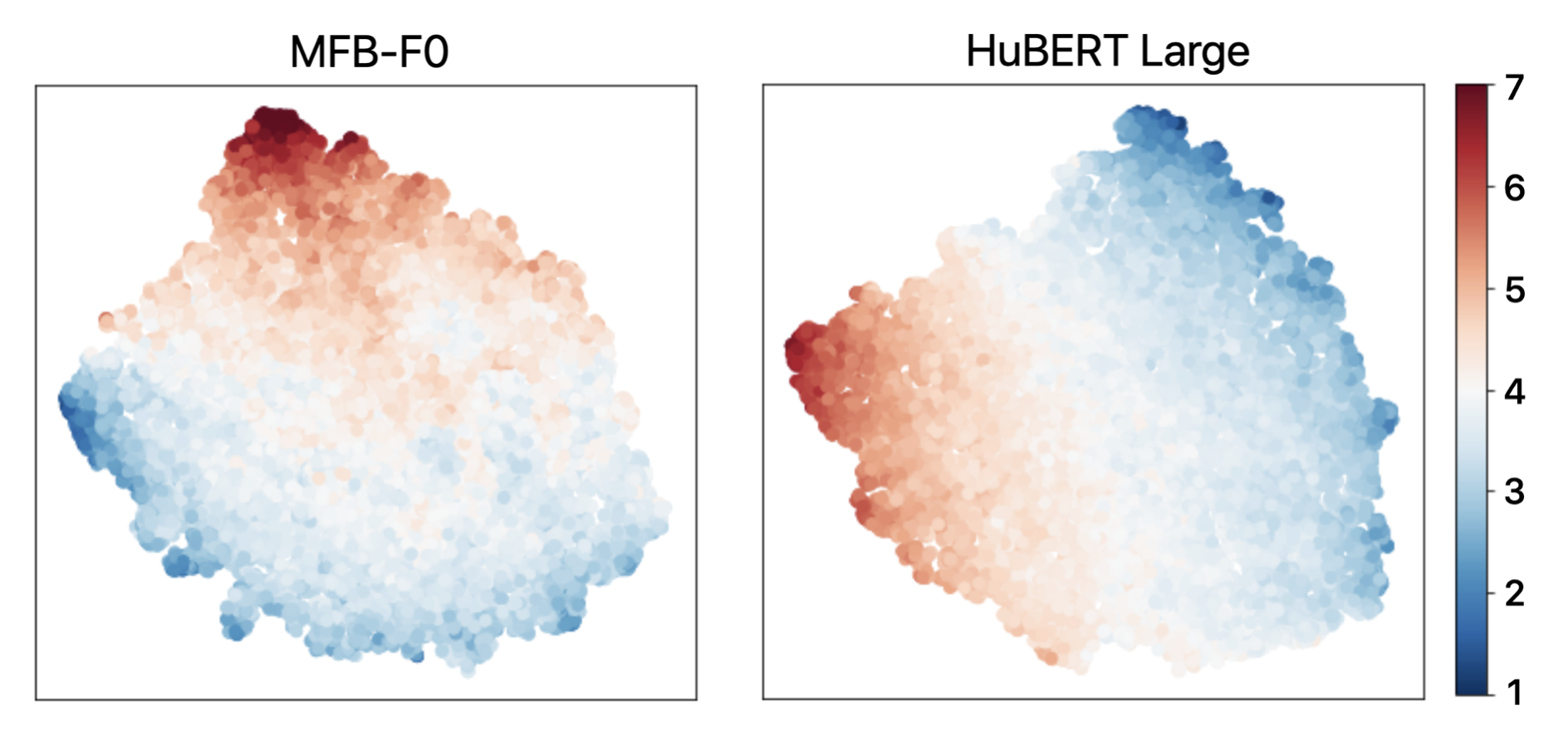}}
\end{minipage}
\caption{t-SNE plot of embeddings obtained from the baseline $MFB+F_0$, and HuBERT embedding trained TCGRU models, colored w.r.t valence score}
\label{fig:fig3}
\end{figure}

\begin{figure}[!ht]
\begin{minipage}[b]{1.0\linewidth}
  \centering
  \centerline{\includegraphics[width=8cm]{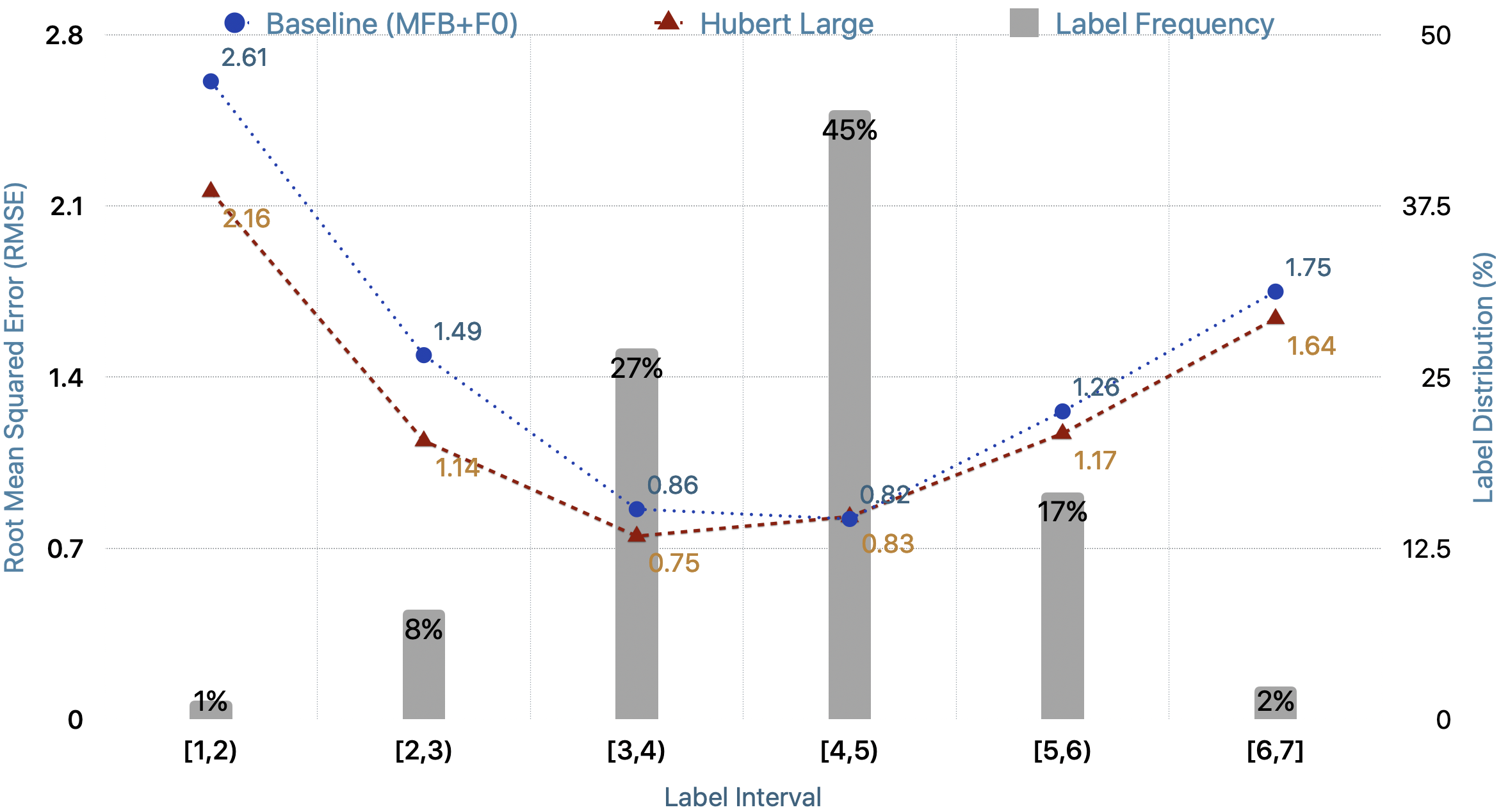}}
\end{minipage}
\caption{Comparing the baseline $MFB+F_0$ trained model with the HuBERT embedding trained model, using root-mean-squared error between the estimate and the target valence scores for different valence label intervals}
\label{fig:fig4}
\end{figure}

We observed a significant ($ \approx 79\%$ relative) performance improvement in valence estimation using the pre-trained model embeddings, compared to $MFB+F_0$ baseline. The t-SNE plots shown in Figure \ref{fig:fig3} demonstrate how the pre-trained model embeddings help better separate the data based on valence scores. Note that the MSP data are heavily skewed to the neutral emotions, and have far less data for high and low valence conditions. We observed that the pre-trained model embeddings help the model to learn edge cases better (compared to the baseline model) as shown in Figure \ref{fig:fig4}. 

\vspace{-1mm}
\begin{table}[th]
  \caption{Dimensional emotion estimation $CCC$ from GRU models trained with pre-trained model embeddings}
  \centering
  \label{tab:table2}
  \begin{tabular}{lccc}
    \toprule
    \textbf{Features}      & act  & val  & dom                \\
    \midrule
    \textbf{MSP-eval\_1.3} \\
    WAV2VEC2.0 BASE             & 0.75  & 0.37  & 0.66  \\
    WAV2VEC2.0 BASE 100H ASR    & 0.72  & 0.51  & 0.60  \\
    WAV2VEC2.0 LARGE 100H ASR      & 0.75  & 0.46  & 0.67  \\
    HuBERT BASE                 & 0.75  & 0.50  & 0.66  \\
    HuBERT LARGE    & \textbf{0.77}  & \textbf{0.53}  & \textbf{0.68}   \\
    HuBERT LARGE 960H ASR        & 0.72  & \textbf{0.53}  & 0.64     \\
    %BERT            & 0.28  & 0.51  & 0.25  \\ 
    \midrule
    \textbf{MSP-eval\_1.6} \\
    WAV2VEC2.0 BASE             & 0.73  & 0.37  & 0.64  \\
    WAV2VEC2.0 BASE 100H ASR    & 0.69  & 0.50  & 0.59  \\
    WAV2VEC2.0 LARGE 100H ASR   & 0.73  & 0.46  & 0.65  \\
    HuBERT BASE                 & 0.73  & 0.50  & 0.64  \\
    HuBERT LARGE   & \textbf{0.74}  & \textbf{0.52}  & \textbf{0.66} \\
    HuBERT LARGE 960H ASR     & 0.70  & 0.51  & 0.62  \\
    %BERT            & 0.27  & 0.49  & 0.23  \\
    \bottomrule
  \end{tabular}
\end{table}
\vspace{-4mm}
\begin{table}[th]
  \caption{Dimensional emotion estimation $CCC$ from TCGRU models trained with HuBERT Large, HuBERT Large 960H embeddings and their combination}
  \centering
  \label{tab:table3}
  \begin{tabular}{lccc}
    \toprule
    \textbf{Features}   & act  & val  & dom                \\
    \midrule
    \textbf{MSP-eval\_1.3} \\
    HuBERT LARGE    & 0.76 & 0.54 & 0.68\\
    HuBERT LARGE 960H  & 0.73  & 0.55  & 0.64     \\
    HuBERT LARGE & & & \\
    + HuBERT LARGE 960H  & \textbf{0.78} & \textbf{0.59}  & \textbf{0.70} \\
 \midrule
    \textbf{MSP-eval\_1.6} \\
    HuBERT LARGE & 0.74 & 0.53 & 0.66 \\
    HuBERT LARGE 960H & 0.70  & 0.53  & 0.62  \\
    HuBERT LARGE & & & \\
    + HuBERT LARGE 960H  & \textbf{0.75} & \textbf{0.57}  & \textbf{0.67} \\
    \bottomrule
  \end{tabular}
\end{table}
\vspace{-4mm}
\begin{table}[ht]
%\small
\centering
\caption{Dimensional emotion estimation $CCC$ from TCGRU models trained with baseline ${MFB+F_0}$ features before and after distillation}
\label{tab:table4}
  \begin{tabular}{lcccccc}
    \toprule
    \multirow{2}{*}{System} &
      \multicolumn{3}{c}{\textbf{MSP-eval\_1.3}} &
      \multicolumn{3}{c}{\textbf{MSP-eval\_1.6}} \\
      & {act} & {val} & {dom} & {act} & {val} & {dom} \\
      \midrule
    -distillation & 0.73 & 0.33 & 0.66 & 0.71 & 0.33 & 0.64  \\
    +distillation & \textbf{0.75} & \textbf{0.37} & \textbf{0.67} & \textbf{0.73} & \textbf{0.37} & \textbf{\textbf{0.65}} \\
    \bottomrule
  \end{tabular}
\end{table}
\vspace{-2mm}
\vspace{1mm}

\section{Conclusions}
In this work we investigated estimation of dimensional emotions (activation, valence and dominance) from speech. We observed a significant improvement (79\% relative) in the performance of valence estimation using pre-trained model embeddings compared to low-level acoustic features (MFBs). The relative improvement in activation and dominance was minor (${\approx 6\%}$ and ${\approx 5\%}$ respectively) from the pre-trained model embeddings as compared to the $MFB+F_0$ features. We observed that having a time-convolutional input layer and a multi-task objective function (that uses discrete emotion detection as an auxiliary task), was beneficial for improving the overall model performance. Furthermore, we explored whether knowledge can be distilled from high-dimensional embedding based models into a low-level feature based models, and observed a relative improvement in $CCC$ of ${\approx 12\%}$ for valence estimation over the baseline MFB trained models. This work shows that while, embedding based models may be complex for real world use cases, they can be used as a teacher network to improve the performance of simpler models that are $19\times$ smaller. 

We report a new state-of-the-art "text-free" acoustic-only dimensional emotion estimation $CCC$ values on two MSP-Podcast evaluation sets. We observed a significant improvement in performance from embedding fusion, which may indicate that such embeddings are offering complementary information that the speech emotion model is benefiting from. Future work should explore embedding features generated from a diverse set of pre-trained models and ways to reduce domain mismatch when adapting those features to target dataset.

\bibliographystyle{IEEEtran}

\bibliography{mybib}

% \begin{thebibliography}{9}
% \bibitem[1]{Davis80-COP}
%   S.\ B.\ Davis and P.\ Mermelstein,
%   ``Comparison of parametric representation for monosyllabic word recognition in continuously spoken sentences,''
%   \textit{IEEE Transactions on Acoustics, Speech and Signal Processing}, vol.~28, no.~4, pp.~357--366, 1980.
% \bibitem[2]{Rabiner89-ATO}
%   L.\ R.\ Rabiner,
%   ``A tutorial on hidden Markov models and selected applications in speech recognition,''
%   \textit{Proceedings of the IEEE}, vol.~77, no.~2, pp.~257-286, 1989.
% \bibitem[3]{Hastie09-TEO}
%   T.\ Hastie, R.\ Tibshirani, and J.\ Friedman,
%   \textit{The Elements of Statistical Learning -- Data Mining, Inference, and Prediction}.
%   New York: Springer, 2009.
% \bibitem[4]{YourName17-XXX}
%   F.\ Lastname1, F.\ Lastname2, and F.\ Lastname3,
%   ``Title of your INTERSPEECH 2022 publication,''
%   in \textit{Interspeech 2022 -- 23\textsuperscript{rd} Annual Conference of the International Speech Communication Association, September 18-22, Incheon, Korea, Proceedings, Proceedings}, 2022, pp.~100--104.
% \end{thebibliography}

\end{document}